%% file: SIL.tex
\documentclass{article}

\usepackage{arxiv}

\usepackage[utf8]{inputenc} 
\usepackage[T1]{fontenc}    
\usepackage{hyperref}       
\usepackage{url}            
\usepackage{booktabs}       
\usepackage{amsfonts}       
\usepackage{nicefrac}       
\usepackage{microtype}      
\usepackage{lipsum}		
\usepackage{graphicx}
\usepackage{natbib}
\usepackage{doi}

\usepackage{amsmath,amssymb,amsthm}
\usepackage[ruled,vlined,linesnumbered]{algorithm2e}
\usepackage{color}
\usepackage{multirow}
\usepackage{enumerate}
\usepackage{setspace}

\input{MyShort}

\newcommand{\mo}[1]{\left\| #1 \right\|_2}

\newcommand{\moo}[1]{\left\| #1 \right\|_{\tau}}

\title{Integrative Learning of Structured High-Dimensional Data from Multiple Datasets}

\author{ Changgee Chang\thanks{Corresponding authors}\\
	Department of Biostatistics, Epidemiology, and Informatics\\
	University of Pennsylvania\\
	Philadelphia, PA 19104 \\
	\texttt{changgee@pennmedicine.upenn.edu} \\
	\And
	Zongyu Dai \\
	Department of Biostatistics, Epidemiology, and Informatics\\
	University of Pennsylvania\\
	Philadelphia, PA 19104 \\
	\texttt{daizy@sas.upenn.edu} \\
	\And
	Jihwan Oh \\
	Department of Biostatistics, Epidemiology, and Informatics\\
	University of Pennsylvania\\
	Philadelphia, PA 19104 \\
	\texttt{jihwan05@pennmedicine.upenn.edu} \\
	\And
	Qi Long\footnotemark[1] \\
	Department of Biostatistics, Epidemiology, and Informatics\\
	University of Pennsylvania\\
	Philadelphia, PA 19104 \\
	\texttt{qlong@upenn.edu} \\
}

\renewcommand{\shorttitle}{CEDAR}

\hypersetup{
pdftitle={A template for the arxiv style},
pdfsubject={q-bio.NC, q-bio.QM},
pdfauthor={Changgee Chang, Zongyu Dai, Jihwan Oh, Qi Long},
pdfkeywords={First keyword, Second keyword, More},
}

\begin{document}
\maketitle

\begin{abstract}
Integrative learning of multiple datasets has the potential to mitigate the challenge of small $n$ and large $p$ that is often encountered in analysis of big biomedical data such as genomics data. Detection of weak yet important signals can be enhanced by jointly selecting features for all datasets. However, the set of important features may not always be the same across all datasets. Although some existing integrative learning methods allow heterogeneous sparsity structure where a subset of datasets can have zero coefficients for some selected features, they tend to yield reduced efficiency, reinstating the problem of losing weak important signals. We propose a new integrative learning approach which can not only aggregate important signals well in homogeneous sparsity structure, but also substantially alleviate the problem of losing weak important signals in heterogeneous sparsity structure. Our approach exploits a priori known graphical structure of features and encourages joint selection of features that are connected in the graph. Integrating such prior information over multiple datasets enhances the power, while also accounting for the heterogeneity across datasets. Theoretical properties of the proposed method are investigated. We also demonstrate the limitations of existing approaches and the superiority of our method using a simulation study and analysis of gene expression data from ADNI.
\end{abstract}


\section{Introduction}
\label{sec:introduction}

Massive amounts of high-throughput -omics data that have been generated in recent studies offer great promises in deepening our understanding molecular underpinning and mechanisms for complex diseases such as Alzheimer's disease and cancer. At the same time, they still present significant analytical challenges as the sample size in a single study is often small to moderate. There is a large body of literature on regularized regression models for the analysis of high-dimensional data in the setting where the number of variables is larger than the sample size. While many of these methods have appealing asymptotic properties, there is a growing recognition that their performance in practice is often unsatisfactory when the sample size is small and the signal-to-noise ratio is small.  A number of approaches have been proposed to mitigate this limitation of regularized regressions, particularly for the analysis of genomics data. 

One popular approach is to incorporate prior knowledge on high-dimensional predictors such as gene regulatory pathways and co-expression networks that are represented by graphs and can be obtained from public or commercial databases such as Kyoto Encyclopedia of Genes and Genomes (KEGG, \citet{KEGG2017}) and Gene Ontology \citep{Ashburner2000}.
The knowledge-guided approach for structured data whose variables lie on a graph \citep{Li2010} has been adopted in supervised learning such as regression \citep{Li2008,Pan2010,Yu2016,Chang2018EMSHS} and in unsupervised learning \citep{li2020bayesian,liu2017novel}, through carefully designed  penalty functions in a frequentist framework or prior specifications in a Bayesian framework.
The rationale behind incorporating the graphical structure of features into supervised learning is the fact that phenotypic biomarkers are often manifested as a result of interaction between a group of genes (pathway).
It is typically not the case that the important features are unrelated. Rather, one or more groups of closely related genes have the predictive power jointly.
Therefore, the graphical information can be integrated by encouraging the group-wise selection of the model coefficients.
For example, \citet{Li2008,Pan2010} propose network-based penalties which encourage joint selection of the predictors that are connected in the graph. \citet{Li2010,Stingo2011} use a Markov random field (MRF) prior combined with a spike and slab prior to encourage selection of connected predictors. More recently, \citet{Chang2018EMSHS} propose a structured shrinkage prior which mitigates some issues associated with prior Bayesian methods.
While all the aforementioned methods use the predictor graph in an edge-by-edge manner and encourage selection of adjacent nodes, \cite{Yu2016} proposed a method that uses the predictor graph in a node-by-node manner and encourage selection of the neighborhood group of each predictor.
These knowledge-guided statistical learning methods have shown improved prediction accuracy and improved power for detecting weak yet important signals in finite samples and they encourage selection of pathways rather than individual features, leading to biologically more meaningful and interpretable results \citep{Zhao2019JCO}.

Another useful approach for mitigating the small sample size problem is integrative learning of multiple datasets that contain the same set of variables, also known as horizontally partitioned data.
Multiple datasets are broadly defined as being collected from multiple studies/sites, from multiple sex/racial groups, or from multiple related disease groups. One key advantage of integrative learning in regression is that it improves the power for detecting important predictors that are shared across the datasets \citep{Zhao2015}.    \citet{Ma2011} proposed an integrative analysis approach that assumes the same sparsity structure of the regression coefficients across all datasets but allows for different effect sizes. The homogeneous sparsity assumption, however, can be overly restrictive in some applications. This assumption is relaxed in subsequent work by, among others, \citet{Li2014,Liu2014,Liu2014Composite,Huang2017} which allow for heterogeneity in sparsity structure across multiple datasets. 
However, these existing integrative learning methods do not account for important graph information for structured predictors such as genomics data, which has the potential to further improve the power of detecting weak yet important signals. Thus the existing heterogeneity models, as they allow the coefficient of a selected feature in some datasets to become zero, may miss such weak, yet important signals, weakening the power of integrative learning.
To the best of our knowledge, there has been little work on incorporating graph information into integrative learning except for \cite{Liu2013}, and their approach relies on the assumption of homogeneous sparsity across all datasets, which may be unrealistic in many applications.
For example, when performing integrative learning of datasets from populations at different stages of a disease, the set of important predictors may vary across these datasets.

To address this gap, we propose a novel integrative learning approach, called Structured Integrative Learning (SIL), which enables incorporation of structural information such as graphical knowledge on predictors.
The key idea underlying SIL is that if a group of features as defined by pathways/networks are important in one dataset, they are likely to be important for the other datasets as well. Our approach is designed to select `groups of features' jointly for all datasets rather than selecting `individual features' jointly. 
As such, it is expected to further improve the power of detecting weak, yet important signals.
Our proposed method can accommodate both homogeneous and heterogeneous sparsity structure, and in particular our method is theoretically justifiable.
We show the oracle inequalities that provide the upper bounds of estimation and prediction errors in a non-asymptotic manner.
We also investigate the conditions for the oracle property to hold in the setting where both the number of datasets and the number of predictors diverge.
Another contribution of our work is to develop an iterative shrinkage-thresholding algorithm \citep{Beck2009,Gong2013} that fits our model, which is much more scalable than the (sub)gradient descent algorithm that has typically been used in prior work for integrative learning.
We show that the proximal operators associated with our regularizers have analytic solutions and can be evaluated very efficiently.  

We note that \citet{Chang2020GRIA} has presented the intermediate results of our research on the proposed method.
Compared to the earlier version, this work includes several significant improvements as follows.
The current work presents a more general penalty formulation of which the penalty in the prior version can be viewed as a special case.
Theoretical properties of the general penalty are rigorously investigated, while the earlier version includes no theoretical result. 
We include new regularizers that are based on log-sum penalty and the efficient algorithms for them. The regularizers in the prior paper are still included and compared in the simulation and data analysis studies.
In the simulation study, we compare performance using both homogeneity data and heterogeneity data, while the earlier version only uses heterogeneity data. Moreover, we perform a sensitivity analysis which investigates robustness of our method against inaccurate and/or incomplete graphical information, while no sensitivity analysis in the earlier work.
Finally, we include a pathway enrichment analysis in data analysis, which demonstrates our method yields outcomes that can be more interpretable and biological meaningful, while the prior work included no pathway enrichment analysis.

The remainder of this article is organized as follows.  We describe the problem of interest and present our proposed method in Section \ref{sec:SIL}, and then present the numerical algorithm in Section \ref{sec:algorithm}.
In Section \ref{sec:theory}, we present the theoretical properties of the proposed method. In Section \ref{sec:simulation}, we conduct simulation studies to evaluate the performance of our approach in comparison with several existing methods. In Section \ref{sec:application}, we further illustrate the strengths of the proposed method through analysis of real data from the Alzheimer's Disease Neuroimaging Initiative (ADNI). We conclude the paper with some discussion remarks in Section \ref{sec:discussion}.

\section{Method}
\label{sec:SIL}
\subsection{Background}
\label{sec:background}
To fix ideas, consider fitting linear regression model using data from $M$ datasets.
In the $m$-th dataset, we have an $n_m \times p$ predictor matrix $X^m$ and an $n_m \times 1$ response vector $\mb y^m$, where $n_m$ is the sample size of the $m$-th dataset and $p$ is the number of predictors.
Let $N = \sum_m n_m$ be the total sample size.
The model of interest is the linear model
\begin{eqnarray}
\mb y^m = X^m \bs \beta^{0m} + \mb e^m, \quad m=1,\dots,M,\label{LM}
\end{eqnarray}
where $\bs \beta^{0m} = (\beta_1^{0m}, \dots,\beta_p^{0m})^T$ is the $p \times 1$ true coefficient vector and $\mb e^m \sim \mc N(0,\sigma^2 I)$ is the $n_m \times 1$ error vector for the $m$-th dataset.
The regularized least square loss function is generally given by
\begin{align*}
\mc L(B) =  \sum_{m=1}^M L_m(\bs \beta^m) + P(B),
\end{align*}
where $B = \left[ \begin{matrix} \bs \beta^1 & \cdots & \bs \beta^M \end{matrix} \right]$ is the $p \times M$ coefficient matrix, $P(B)$ is a penalty on $B$, and
\begin{align*}
L_m(\bs \beta^m) =  \frac{1}{2n_m} \left\| \mb y^m - X^m \bs \beta^m \right\|_2^2.
\end{align*}

The most general estimator would allow all $\bs \beta^m$ to be different and use a separable penalty $P(B) = \sum_{m=1}^M P_m(\bs \beta^m)$.
This is equivalent to independently minimizing
\begin{align*}
\mc L_m(\bs \beta^m) =  L_m(\bs \beta^m) +  P_m(\bs \beta^m),
\end{align*}
which is equivalent to analyze each dataset separately.
We call this model the fully heterogeneous model.
On the other hand, the least general estimator would assume all coefficients to be the same across all datasets; $\bs \beta^m \equiv \bs \beta$. 
This fully homogeneous model is equivalent to merging all datasets with each data point weighted by the reciprocal of the size of the dataset it belongs to.
The weights prevent the large dataset from dominating the loss function and keep the coefficients from leaning favorably only to the large dataset.


Obviously, the full homogeneity assumption can be overly restrictive.
Each dataset often has its own characteristics, and the association between the outcome and the predictors can be different.
Ignoring the difference can lead to poor or suboptimal performance in estimation and prediction.
On the other hand, the fully heterogeneous model fails to borrow information across datasets, and the regression model for each dataset can suffer the curse of dimensionality. The motivation of integrative learning is to aggregate common information from multiple datasets while accounting for heterogeneity across these datasets.

\subsection{Structured Integrative Learning}

In this work, we focus on the case where the graph information for predictors is the same across the $M$ datasets.
Denote by $G = \langle V,E \rangle$ the graph on predictors $X$ where $V = \{1,\dots,p\}$ is the set of features and $E$ is the set of edges between the features.
Let $A=[a_{jk}]$ be the adjacency matrix associated with $G$ and let $\mc A_j = \{k:a_{jk}=1\} \cup \{j\}$ be the neighborhood of the $j$-th feature including itself.
Let $e = |E|$ be the number of edges in $G$ and $a_j = |\mc A_j|$ be the number of members in $\mc A_j$.
The graphical information on features often represents the partial correlation structure of the features.
That is, the presence of an edge between features $j$ and $k$ implies the $(j,k)$ entry of the feature precision matrix is nonzero, while an absence of edge means a zero entry.
In analysis of genomics data, such graphs often represent gene regulatory pathways or co-expression networks which can be obtained from existing databases such as KEGG \citep{KEGG2017}.

For Model~\eqref{LM}, note that we have $\bs \beta^m = \Omega^m \mb c^m$ where $\Omega^m = n_m E(X^{mT} X^m)^{-1}$ and $\mb c^m = \frac{1}{n_m} E(X^{mT} \mb y^m)$.
This yields
\begin{align} \label{eqn:motif}
\bs \beta^m = \sum_{j=1}^p c_j^m \bs\omega_j^m,
\end{align}
where $\bs\omega_j^m$ is the $j$-th column of $\Omega^m$.
Since the absence of an edge implies a zero partial correlation, we have $\supp(\bs\omega_j^m) = \mc A_j$.
Following the observation from \citet{Yu2016}, either $\bs \beta_{\mc A_j}^m \equiv (\beta_k^m)_{k \in \mc A_j}$ can have nonzero values (if $c_j^m \neq 0$), or there will be no contribution from the $j$-th group to the effect size $\bs \beta^m$ (if $c_j^m = 0$).
The key premise of our work is that if a group $\mc A_j$ is important for one dataset, it is likely to be important for other datasets as well.
To encourage joint selection of the feature groups $\mc A_j$ across all datasets, we propose the following penalty.
\begin{align} \label{eqn:SIL}
    P(B) = \lambda \|B\|_\tau, \quad \|B\|_\tau \equiv \inf_{B = \sum_j \Gamma_j, \supp(\bs\gamma_j^m) = \mc A_j} \sum_{j=1}^p \tau_j \rho (\Gamma_j),
\end{align}
where $\Gamma_j = \begin{bmatrix} \bs\gamma_j^1 & \cdots & \bs\gamma_j^M \end{bmatrix}$ in an arbitrary $p \times M$ matrix. 

Note that the standard group lasso penalty such as $P(B) = \lambda \sum_{j=1}^p \tau_j \|B_{\mc A_j}\|_F$ where $B_{\mc A_j} = \begin{bmatrix} \bs \beta_{\mc A_j}^1 & \cdots & \bs \beta_{\mc A_j}^M \end{bmatrix}$ and $\| \cdot \|_F$ is the Frobenius norm has an undesirable characteristic.
Since $\mc A_j$ are overlapping in general, it yields $\beta_j^m \neq 0$ if and only if all groups including $j$ are selected.
In other words, gene $j$ can be selected if and only if all groups (pathways) including $j$ are important, which is not plausible in biology.
On the other hand, the proposed penalty \eqref{eqn:SIL} introduces latent coefficients $\Gamma_j$ and can yield $\beta_j^m \neq 0$ if at least one group that contains $j$ is selected.
That is, if a pathway is important, then all genes in the pathway can become important.
This is a required property to be consistent with \eqref{eqn:motif}.
More details about the latent group lasso penalty can be found in \citet{Jacob2009}.
It is easy to see that applying the proposed penalty is equivalent to replacing $\bs \beta^m$ with $\sum_j \bs\gamma_j^m$ in the loss function and enforcing a penalty on $\bs\Gamma = (\Gamma_1,\dots,\Gamma_p)$ as follows.
\begin{align*}
L(\bs\Gamma) = \sum_{m=1}^M \frac{1}{2n_m} \left\| \mb y^m - X^m \sum_{j=1}^p \bs\gamma_j^m \right\|_2^2, \quad P(\bs\Gamma) = \lambda \sum_{j=1}^p \tau_j \rho(\Gamma_j), \quad \supp(\bs\gamma_j^m) = \mc A_j.
\end{align*}
Minimizing $L(\bs\Gamma)+P(\bs\Gamma)$ will yield the same solution as minimizing $L(B)+P(B)$ with $\whbs \beta^m = \sum_j \whbs\gamma_j^m$.
The efficient algorithms for selected penalties will be presented in Section \ref{sec:algorithm}.

We propose the core penalty function $\rho(\Gamma_j)$ has the following form.
\begin{align} \label{eqn:core}
    \rho(\Gamma_j) = \rho_1 \circ \rho_2 (\Gamma_j).
\end{align}
The outer penalty $\rho_1: \bb R_+ \rightarrow \bb R_+$ determines the sparsity and unbiasedness levels and the positive halves of all popular concave penalty functions can be used; for example, lasso \citep{Tibshirani1996}, SCAD \citep{Fan2001}, MCP \citep{Zhang2010}, and log-sum \citep{Candes2008,Armagan2013}.
The inner penalty $\rho_2: \bb R^{p\times M} \rightarrow \bb R_+$ determines how the penalties of the columns $\bs\gamma_j^m$ are combined, and yields different type of models.
The choice of $\rho_2(\Gamma) = \|\Gamma\|_F$ selects all components in an all-in-or-all-out fashion and leads to the homogeneity model, but the choice of $\rho_2(\Gamma) = \|\Gamma^T\|_{2,1}$ ($L_{2,1}$ norm) allows each $\bs\gamma_j^m$ to become zero and leads to the heterogeneity model.
The combination of both $\rho_2(\Gamma) = \alpha \|\Gamma\|_F + (1-\alpha) \|\Gamma^T\|_{2,1}$ can also be considered.

For example, we can choose the log-sum penalty for $\rho_1$.
\begin{align} \label{eqn:logsum}
    \rho_{LS}(x) = \eta \log (1+x/\eta).
\end{align}
Then, depending on the choice for $\rho_2$, we can have two types of penalties
\begin{align} \label{eqn:LSpen}
    P_{LS_1}(\bs\Gamma) = \lambda \sum_{j=1}^p \tau_j \rho_{LS} ( \|\Gamma_j\|_F ), \quad P_{LS_2}(\bs\Gamma) = \lambda \sum_{j=1}^p \tau_j \rho_{LS} ( \|\Gamma_j^T\|_{2,1} ).
\end{align}
In both choices, the entries of $\bs\gamma_j^m$ becomes zero or nonzero in an all-in-or-all-out fashion.
The weights $\tau_j$ should take into account the size of the group $\mc A_j$ and thus is recommended to have the form $\tau_j = \sqrt{a_j} d_j$ for some $d_j>0$.
We can have homogeneous penalty ($d_j=1$), or we can reflect the importance of the group $\mc A_j$ by choosing adaptive penalty $d_j$.
Noting that, for example, the coefficients are proportional to the absolute correlation between the response variable and the predictor $j$ as in \eqref{eqn:motif}, 
$d_j$ can be chosen to be reciprocal to the average of the absolute sample correlations $d_j^{-1} = M^{-1} \left| \sum_m \mb x_j^{mT} \mb e^m / n_m\right|$.

Note that the $L_1$ or $L_2$ norm based penalty discourages to include highly correlated variables/groups simultaneously.
In other words, true signals can be pushed away from the model by its highly correlated neighbors, and it is the case in our model as well.
Due to the construction of $\mc A_j$, some $\mc A_j$'s are often similar or can even be identical.
As a remedy to this issue, we follow the idea of the elastic net \citep{Zou2005}, which adds the ridge penalty.
\begin{align*}
    P_R(B) = \frac{\lambda_R}{2}  \| B \|_F^2.
\end{align*}
The ridge penalty seemingly reduces the correlations between groups and facilitates inclusion of all potentially important signals.
This completes the objective function of our model.
\begin{align*}
    \mc L(B) = L(B) + P_R(B) + P(B),
\end{align*}
or equivalently
\begin{align*}
    \mc L(\bs\Gamma) = L(\bs\Gamma) + P_R(\bs\Gamma) + P(\bs\Gamma), \quad \supp(\bs\gamma_j^m) = \mc A_j,
\end{align*}
where
\begin{align*}
    P_R(\bs\Gamma) = \sum_{j=1}^p \frac{\lambda_R a_j}{2} \| \Gamma_j \|_F^2.
\end{align*}

Many existing penalties can be viewed as a special case of ours.
If we have only one dataset and we choose $\rho_1(x) = x$ and $\rho_2(\Gamma) = \|\Gamma\|_F$, then our method simplifies to the sparse regression incorporating graphical structure among predictors \citep{Yu2016}.
If no edge exists or no graphical information is available, we have $\mc A_j = \{j\}$ for all $j$.
In this case, if we choose the minimax concave penalty for $\rho_1$, our method reduces to the methods introduced in \citet{Liu2014Composite}.
If we choose $\rho_1(x)=x$ and the Frobenius norm for $\rho_2$, our model simplifies to the integrative analysis model with the group lasso penalty proposed by \citet{Ma2011}.

\section{Algorithm}
\label{sec:algorithm}

In this section, we present an efficient algorithm to fit our model with the log-sum penalties defined in \eqref{eqn:LSpen}, which will enhance its usefulness in analysis of high-dimensional data such as genomics data.
We also consider two other penalties in this work. Instead of \eqref{eqn:logsum}, we can use the MCP penalty \citep{Zhang2010} for $\rho_1$.
\begin{align} \label{eqn:MCP}
\rho_{MCP}(x) = \int_0^x (1-u/(\lambda\eta))_+ du.
\end{align}
Or, we can have the convex penalty
\begin{align} \label{eqn:Lasso}
    \rho_1(x) = x, \quad \rho_2(\Gamma) = \alpha \|\Gamma\|_F + (1-\alpha) \|\Gamma^T\|_{2,1}.
\end{align}
The algorithms for \eqref{eqn:MCP} and \eqref{eqn:Lasso} can be found in \citet{Chang2020GRIA}.

Let $\bs\delta^m = (\bs\delta_1^{mT}, \dots, \bs\delta_p^{mT})^T$ and $\Delta_j = \begin{bmatrix} \bs\delta_j^1 & \cdots & \bs\delta_j^M \end{bmatrix}$ where $\bs\delta_j^m = \bs\gamma_{\mc A_j,j}^m$ is the vector of unconstrained coefficients in $\bs\gamma_j^m$.
Let $Z_j^m$ be the submatrix of $X^m$ including the columns corresponding to $\mc A_j$ and let $Z^m = \begin{bmatrix} Z_1^m & \cdots & Z_p^m \end{bmatrix}$.
Denoting $\bs\Delta = (\Delta_1,\dots,\Delta_p)$, our objective function can be decomposed into a differentiable part $L(\bs\Delta)+P_R(\bs\Delta)$ and a non-differentiable part $P_{LS_2}(\bs\Delta)$ or $P_{LS_1}(\bs\Delta)$ where
\begin{align*}
L(\bs\Delta) &= \sum_{m=1}^M \frac{1}{2n_m} \left\| \mb y^m - Z^m \bs\delta^m \right\|_2^2, &\quad P_R(\bs\Delta) &= \sum_{j=1}^p \frac{\lambda_R a_j}{2} \left\| \Delta_j \right\|_F^2,\\
P_{LS_1}(\bs\Delta) &= \sum_{j=1}^p \lambda \eta \tau_j \log ( 1 + \|\Delta_j\|_F/\eta ), &\quad
P_{LS_2}(\bs\Delta) &= \sum_{j=1}^p \lambda \eta \tau_j \log ( 1 + \|\Delta_j^T\|_{2,1} /\eta ) .
\end{align*}

We use the accelerated proximal gradient descent algorithm (FISTA, \citet{Beck2009}) to fit our models.
While the log-sum penalty is not convex, its second derivative is bounded from below and it satisfies the criteria in \citet{Gong2013}.
Propositions \ref{pro:LS} and \ref{pro:SLS} describe how to evaluate the proximal operators for $P_{LS_1}$ and $P_{LS_2}$, respectively.
Let $\widetilde{\bs\Delta}$ be the proximal operator associated with penalty $P(\bs\Delta)$ evaluated at $\bs\Delta$, as defined below.
\begin{align*}
\widetilde{\bs\Delta} \equiv \prox_t(\bs\Delta) \equiv \argmin_{\mb W=(W_1,\dots,W_p)} \left( \frac{1}{2t} \sum_{j=1}^p \| W_j - \Delta_j \|_F^2 + P(\mb W) \right).
\end{align*}


\begin{pro} \label{pro:LS}
For $t < \eta/(\lambda \max_j \tau_j)$, the proximal operator associated with the penalty $P_{LS_1}(\bs\Delta)$ is given by
\begin{align} \label{eqn:proxLS}
\widetilde{\Delta}_j = (1-\lambda t  \tau_j h_j/\|\Delta_j\|_F)_+ \Delta_j, \quad j=1,\dots,p,
\end{align}
where
\begin{align} \label{eqn:hLS}
h_j = \frac{1 + \|\Delta_j\|_F/\eta - \sqrt{(1 + \|\Delta_j\|_F/\eta)^2 -4 \lambda t \tau_j/\eta}}{2\lambda t \tau_j/\eta}.
\end{align}
\end{pro}

\begin{pro} \label{pro:SLS}
For $t < \eta/(\lambda \max_j \tau_j)$, the proximal operator  associated with the penalty $P_{LS_2}(\bs\Delta)$ is given by
\begin{align} \label{eqn:proxSLS}
\widetilde{\bs\delta}_j^m = (1-\lambda t  \tau_j h_j/\|\bs\delta_j^m\|_2)_+ \bs\delta_j^m,
\end{align}
where $h_j$ satisfies
\begin{align} \label{eqn:hSLS}
h_j = \frac{1}{1+\sum_{l=1}^M (\|\bs\delta_j^l\|_2 - \lambda t  \tau_j h_j)_+/\eta}.
\end{align}
\end{pro}

The proofs for Propositions \ref{pro:LS} and \ref{pro:SLS} are included in Web Appendix A.
Note that \eqref{eqn:hSLS} is a piecewise quadratic equation in $h_j$ whose analytic solution can be easily obtained as follows.
Let $\xi_j^m = \|\bs\delta_j^m\|_2/( \lambda t  \tau_j)$.
Equation \eqref{eqn:hSLS} can be rewritten as
\begin{align} \label{eqn:hSLS2}
h_j = \frac{1}{1+\lambda t  \tau_j \sum_{l=1}^M (\xi_j^l - h_j)_+/\eta}.
\end{align}
Sort $\xi_j^1, \dots, \xi_j^M$ in ascending order ($\mc O(M\log M)$) and assume, for simplicity, that
\begin{align*}
0 = \xi_j^0 \le \xi_j^1 \le \cdots \le \xi_j^K < 1 \le \cdots \le \xi_j^M,
\end{align*}
for some $K\le M$.
First, note that $h_j=1$ if and only if $\xi_j^M \le 1$.
Suppose $\xi_j^M > 1$ and $h_j \in [\xi_j^{k-1},\xi_j^k)$. From \eqref{eqn:hSLS2}, we have the candidate solution $h_j^k$ as follows.
\begin{align*}
    h_j^k = \frac{1 + \lambda t \tau_j \sum_{l=k}^M \xi_j^l /\eta - \sqrt{(1 + \lambda t \tau_j \sum_{l=k}^M \xi_j^l/\eta)^2 -4 \lambda t \tau_j(M-k+1)/\eta}}{2\lambda t \tau_j (M-k+1)/\eta}.
\end{align*}
If $h_j^k \in  [\xi_j^{k-1},\xi_j^k)$ for some $k \in \{1,\dots,K\}$, it is indeed the solution for \eqref{eqn:hSLS}.
Otherwise, $h_j^{K+1} \in [\xi_j^K,1)$ is the solution for \eqref{eqn:hSLS}.

The algorithm uses the standard accelerated proximal gradient descent algorithm with the backtracking line search.
Each iteration requires $\mc O(pN+Me)$ for $P_{LS_1}$ and $\mc O(pN+Me+pM\log M)$ for $P_{LS_2}$.
We have also investigated the non-accelerated proximal gradient descent algorithm and found that the accelerated version has a substantial advantage when the sample size is small and the ridge penalty $\lambda_R$ is 0 or close to 0.

\section{Theoretical Properties}
\label{sec:theory}

In this section, we study the theoretical properties of the proposed method.
The main goal is to provide the conditions for which the oracle inequality and the oracle property hold in the context of integrative analysis.
Although the Theorem statements and the proofs may look similar to those in \citet{Yu2016}, the implications apply to the analysis of a large number of datasets.
Also, note that the result of $\sqrt{n}$-consistency presented here (Theorem \ref{thm:normality}) is more general than that of \citet{Yu2016}, as the oracle property therein is discussed with fixed $p$ only.

Let $J_0^m = \{j:\beta_j^{0m} \neq 0\}$ be the set of important variables of the $m$th dataset and $J_0 = \bigcup_{m=1}^M J_0^m$ be the union of all important variables.
Define $s_0 = |J_0|$ as the number of all important variables.
Let $J_1 = \{j \in J_0:\mc A_j \subset J_0\}$ be the set of groups which contain important features only, $J_2 = \{j :\mc A_j \cap J_0 \neq \emptyset\}$ be the set of groups which contain at least one important gene, and $J_3 = \{j \in J_0^c:\mc A_j \subset J_0^c\}$ be the set of groups which contain unimportant features only, and let $s_1 = |J_1|$, $s_2 = |J_2|$, and $s_3 = |J_3|$.
Focusing on general penalties with $\rho(\cdot) \ge \|\cdot\|_F$, we first present the oracle inequalities under homogeneous penalty weights $d_j=1$, i.e., $\tau_j = \sqrt{a_j}$ for $j=1,\dots,p$. These are non-asymptotic finite sample properties which account for a diverging number of datasets and predictors.
Then, we discuss the model selection consistency and the asymptotic normality under adaptive penalty weights $d_j$.
To this end, define $d^* = \max_{j\in J_1} d_j$ and $d_* = \min_{j\in J_1^c} d_j$.
Noting that the ridge penalty is not required for the oracle properties to hold and only needed to be small enough, we set $\lambda_R=0$ in this section.

For simplicity, we assume $n_m=n$ for $m=1,\dots,M$ and thus $N=Mn$, and define
\begin{align*}
\mb y = \frac{1}{\sqrt{n}} \begin{bmatrix}
\mb y^1\\
\vdots\\
\mb y^M
\end{bmatrix}, \quad
\mb X = \frac{1}{\sqrt{n}} \diag(X^1,\dots,X^M), \quad
\bs \beta = \vect (B), \quad
\mb e = \frac{1}{\sqrt{n}} \begin{bmatrix}
\mb e^1 \\
\vdots \\
\mb e^M
\end{bmatrix}.
\end{align*}
We vectorize $\Gamma_j$ in this section, $\bs\gamma_j = \vect(\Gamma_j)$, and the penalty is written as
\begin{align*}
\moo{\bs \beta}  = \min_{\bs \beta = \sum_{j=1}^p {\bs\gamma}_j} \sum_{j=1}^p \tau_j  \rho(\bs\gamma_j), \quad \supp({\bs\gamma}_j^m) = \mc A_j,
\end{align*}
with a little abuse of notation; $\rho(\bs\gamma_j) \equiv \rho(\Gamma_j)$.
Then, the objective function goes as follows.
\begin{align} \label{eqn:obj_beta}
    \frac{1}{2} \mo{\mb y-\mb X \bs \beta}^2 + \lambda \sum_{j=1}^p \moo{\bs \beta}.
\end{align}
Let $\widehat{\bs \beta}$ be the minimizer of \eqref{eqn:obj_beta} and $\widehat{\bs\gamma}_1,\dots,\widehat{\bs\gamma}_p$ be an optimal decomposition of $\widehat{\bs \beta}$.

We present the oracle inequalities for estimation and prediction errors.
Let $Q_1 = \max_{m,j} \|\mb x_j^m\|_2^2/n$ be the largest empirical variance of predictors.
Let $\bs \beta^0 = \vect(B^0)$ be the stacked true regression coefficients and $\bs \beta_{J_0}^0$ be the stacked nonzero true regression coefficients.
Let $\beta_*^0$ be the smallest absolute value of nonzero true coefficients across all datasets.
For $\bs \beta \in \bb R^p$, let $U(\bs \beta) = \{(\bs\gamma_1,\dots,\bs\gamma_p) : \sum_j \bs\gamma_j = \bs\beta,\|\bs\beta\|_\tau = \sum_j \tau_j \rho(\bs\gamma_j), \supp(\bs\gamma_j^m) = \mc A_j\}$ be the set of all optimal decompositions of $\bs \beta$, and $K_\tau(\bs \beta)$ be the number of nonzero $\bs\gamma_j$'s in the optimal decomposition of $\bs \beta$ which has the minimal number of nonzero $\bs\gamma_j$'s, i.e., $K_\tau(\bs \beta) = \min_{\bs\Gamma \in U(\bs \beta)} |\{j: \bs\gamma_j \neq \mb 0 \}|$.
Denote $K_\tau = \sup_{\supp(\bs \beta) \subset  (\bigcup_{j \in J_2} \mc A_j)} K_\tau(\bs \beta)$.
We can check $J_1=J_0$, $J_2=J_0$, $J_3=J_0^c$, and $K_\tau = s_0$ if the graph $G$ has no edge.
We need the following assumptions.
\begin{ass}
The important feature set $J_0$ is covered by $\{\mc A_j: j\in J_1\}$. That is, $\bigcup_{j \in J_1} \mc A_j = J_0$.

\label{ass:cover}
\end{ass}

\begin{ass} \label{ass:Gaussian}
The errors $\mb e^m \sim_{iid} \mc N(0, \sigma^2 I)$ for $m = 1,\dots, M$.
\end{ass}
\begin{ass} \label{ass:restricted}
There exists a constant $\kappa>0$ such that 
\begin{align*}
    \inf_{|J|\le s_2, \bs \beta\in \bb R^p \backslash \{0\}} \inf_{\bs\Gamma \in \mc T_\tau(\bs \beta, J) } \frac{\|{\mb X}\bs \beta\|_2}{\sqrt{\sum_{j\in J} \tau_j^2 \rho(\bs\gamma_j)^2}} \ge \kappa,
\end{align*}
where $\mc T_\tau(\bs \beta, J)$ is the set of all optimal decompositions $\bs\Gamma = (\bs\gamma_1,\dots,\bs\gamma_p)$ of $\bs \beta$ such that $\sum_{j\in J^c} \tau_j \rho(\bs\gamma_j) \le 3 \sum_{j\in J} \tau_j \rho(\bs\gamma_j)$.
\end{ass}
In order to select the correct model, the groups that include any unimportant variable must not be selected and only the groups that have important variables only may be selected.
Assumption \ref{ass:cover} ensures that all important variables are covered by the groups with important variables only. 
Although we assume Gaussian errors in Assumption \ref{ass:Gaussian}, the asymptotic properties presented in this paper hold for any iid mean zero sub-Gaussian errors.
Assumption \ref{ass:restricted} is similar to the restricted eigenvalue condition or the compatibility condition \citep{Bickel2009} which is commonly used for these types of inequalities but has been tailored to our proposed penalty.

\begin{thm} \label{thm:inequality} (Oracle inequalities)
Suppose Assumptions \ref{ass:cover}, \ref{ass:Gaussian}, and \ref{ass:restricted} hold.
Assume $\rho(\bs\gamma)$ is a norm such that $\rho(\bs\gamma) \ge \|\bs\gamma\|_2$.
Let $d_j=1$, i.e., $\tau_j = \sqrt{a_j}$ for $j=1,\dots,p$.
If we choose $\lambda \ge 4 \sigma \sqrt{MQ_1} \sqrt{\frac{A + 2\log(Mp)}{n}}$ for some $A>0$, then the following inequalities hold with probability at least $1-2\exp(-A/2)$.
\begin{align*}
    \mo{{\mb X}(\widehat{\bs \beta}-\bs \beta^0)} \le \frac{4\lambda K_\tau^{1/2}}{ \kappa }, \quad \moo{\widehat{\bs \beta}-\bs \beta^0} \le \frac{16\lambda  K_\tau}{ \kappa^2 }, \quad \mo{\widehat{\bs \beta}-\bs \beta^0} \le \frac{16\lambda  K_\tau}{\kappa^2}.
\end{align*}
\end{thm}

Please see Web Appendix B for proofs.
Note that the results of Theorem \ref{thm:inequality} are general and consistent with the results shown in existing literature.
For example, if $M=1$ and we choose $\rho_1(x)=x$ and $\rho_2(\Gamma) = \| \Gamma \|_F$, we obtain the same results as in \citet{Yu2016}.
If, in addition, there is no edge in the graph, we obtain the results similar to \citet{Bickel2009}.

We now present the oracle property focusing on the homogeneity model $\rho(\bs\gamma) = \|\bs\gamma\|_2$.
The objective function can be written in terms of $\bs\Gamma$ as follows.
\begin{align} \label{eqn:objective}
    \frac{1}{2} \mo{\mb y-\mb X\sum_{j=1}^p\bs\gamma_j}^2 + \lambda \sum_{j=1}^p \tau_j \mo{\bs\gamma_j}, \quad \supp(\bs\gamma_j^m) = \mc A_j.
\end{align}
Let $\widehat{\bs\gamma}_1,\dots,\widehat{\bs\gamma}_p$ be the minimizer of \eqref{eqn:objective} and $\widehat{\bs \beta} = \sum_{j=1}^p \widehat{\bs\gamma}_j$ be the solution. 
Let $\mc R \subset 2^{J_1}$ represent the set of subsets of $\{\mc A_j:j \in J_1\}$ which covers the important variables $J_0$.
That is, $R \in \mc R$ if and only if $\bigcup_{j \in R} \mc A_j = J_0$.
Define $R_0 = \argmin_{R \in \mc R} \sum_{j \in R} \tau_j^2$ and $S_0 = \sum_{j \in R_0} a_j$.
This set is not empty due to Assumption \ref{ass:cover}.
Note that we have $S_0 = s_0$ if the graph $G$ has no edge.
Let $Q_2 >0$ be the smallest eigenvalue of $\mb X_{J_0}^T \mb X_{J_0}$ and let $\xi = \|\mb X_{J_0^c}^T \mb X_{J_0} (\mb X_{J_0}^T \mb X_{J_0})^{-1}\|_\infty$.
In Theorems \ref{thm:model} and \ref{thm:normality}, we present low level conditions required for model selection consistency and asymptotic normality, respectively.
In Corollaries \ref{cor:adap1} and \ref{cor:adap2}, we list conditions for individual parameters required for the oracle property, which will depend on the adaptivity of the penalty weights $d_j$.

\begin{thm} \label{thm:model} (Model selection consistency)
Suppose Assumptions \ref{ass:cover} and \ref{ass:Gaussian} hold.
Consider $\rho(\bs\gamma) = \|\bs\gamma\|_2$.
If
\begin{align} \label{eqn:ModelSelCond}
    \frac{\sqrt{\log (M s_0)}}{\beta_*^0 \sqrt{n Q_2}} + \frac{\lambda \sqrt{S_0} d^*}{Q_2 \beta_*^0} +  \frac{\sqrt{M Q_1 \log M(p-s_0)}}{\lambda d_* \sqrt{n}} + \frac{\max(\xi,1)} d^* \sqrt{M S_0}{d_*} \rightarrow 0,
\end{align}
then we have $\sign(\widehat{\bs \beta}) = \sign(\bs \beta^0)$ with probability tending to 1.
\end{thm}

\begin{rem}
The first two terms in \eqref{eqn:ModelSelCond} control the deviation of the nonzero coefficients from their ground truth.
The last two terms in \eqref{eqn:ModelSelCond} ensure the penalties are large enough to suppress the coefficients of unimportant predictors.
\end{rem}

Our method also possesses the property of asymptotic normality.
However, in order to have $\sqrt{n}$-consistency, we need a stronger condition compared to the model selection consistency.
\begin{thm} (Asymptotic normality) \label{thm:normality}
Assume the conditions in Theorem \ref{thm:model}, and further assume \begin{align} \label{eqn:AsymNormalCond}
    \frac{ \lambda d^* \sqrt{nS_0}}{\sqrt{Q_2}} \rightarrow 0.
\end{align}
Let $v = \bs\alpha^T({\mb X}_{J_0}^{T}{\mb X}_{J_0})^{-1} \bs\alpha$ for any sequence of nonzero vector $\bs\alpha$ with length $M J_0$.
Then, we have
\begin{align*}
    \sqrt{n} \bs\alpha^T (\widehat{\bs \beta}_{J_0} - \bs \beta_{J_0}^0) / \sqrt{v} \rightarrow_d \mc N(0,\sigma^2).
\end{align*}
\end{thm}

We now investigate conditions for individual factors which guarantee the oracle property.
\begin{ass} \label{ass:s0}
    $S_0 \asymp s_0 \asymp n^\alpha$ where $0 \le \alpha <1$.
\end{ass}
\begin{ass} \label{ass:Q}
    $Q_1 \asymp Q_2 \asymp \xi \asymp 1$.
\end{ass}
\begin{ass} \label{ass:beta0}
    $\beta_*^0 \asymp s_0^{-1/2} \asymp n^{-\alpha/2}$.
\end{ass}
\begin{ass} \label{ass:lam_upper}
    $\lambda = o(n^{-(1+\alpha)/2})$ and $d^* \asymp 1$.
\end{ass}
The number of important variables must typically be less than the sample size.
This is also connected in part to the condition on the smallest eigenvalue $Q_2$ of $\mb X_{J_0}^T \mb X_{J_0}$ as in Assumption \ref{ass:Q}.
The predictors can always be standardized, so we can have $Q_1 \asymp 1$ as well.
The assumption $\xi \asymp 1$ is similar to but weaker than the irrepresentable condition \citep{Zhao2006} since the bound needs not be less than 1.
We consider the signal-to-noise ratio fixed at a constant level.
Therefore, $\| \bs \beta_{J_0}^{0m}\|_2 \asymp 1$ and Assumption \ref{ass:beta0} are plausible.
Assumption \ref{ass:lam_upper} sets a penalty cap which limits the bias for important variables caused by the penalty.
The conditions for $M$, $p$ and the lower bound of $\lambda$ depend on the minimum adaptive penalty weights $d_*$ on the unimportant variables.
\begin{cor} \label{cor:adap1}
(Strongly adaptive penalty weights)
Suppose Assumptions \ref{ass:s0}--\ref{ass:lam_upper} hold.
If $d_* \asymp N^{\gamma/2}$ with $\gamma \ge 1$, then the conditions \eqref{eqn:ModelSelCond} and \eqref{eqn:AsymNormalCond} are satisfied if
\begin{align*}
    \log M = o(n^{1-\alpha}), \quad \log p = o(n^{1-\alpha}), \quad \lambda^{-1} = o(n (\log (Mp))^{-1/2}).
\end{align*}
\end{cor}

If the adaptive penalty weights for unimportant variables are chosen at a rate of $\sqrt{N}$ or higher, the number $M$ of datasets our method can accommodate for the oracle property only depends on the number of important variables, and we can have exponentially growing number of datasets with respect to $n$ raised to a certain power.
However, we find that, if the penalty weights are weakly adaptive, which means the minimum adaptive penalty weights for unimportant variables are chosen at a rate lower than $\sqrt{N}$, our method may only accommodate polynomially increasing number of datasets with respect to $n$.

\begin{cor} \label{cor:adap2}
(Weakly adaptive penalty weights)
Suppose Assumptions \ref{ass:s0}--\ref{ass:lam_upper} hold.
If $d_* \asymp N^{\gamma/2}$ with $\alpha < \gamma < 1$, then the conditions \eqref{eqn:ModelSelCond} and \eqref{eqn:AsymNormalCond} are satisfied if
\begin{align*}
    M = o(n^{\frac{\gamma-\alpha}{1-\gamma}}), \quad \log p = o(M^{-(1-\gamma)} n^{\gamma-\alpha}), \quad \lambda^{-1} = o( M^{-\frac{1-\gamma}{2}} n^{\frac{1+\gamma}{2}} (\log (Mp))^{-\frac{1}{2}}).
\end{align*}
\end{cor}

It is worth noting that while the oracle inequality (Theorem \ref{thm:inequality}) holds with the convex penalty and no adaptation ($d_j=1$), the oracle property (Theorems \ref{thm:model} and \ref{thm:normality}) requires an adaptive penalty.
This result is consistent with the behavior of the ordinary lasso regression.
The $L_1$ penalty can achieve the oracle inequality \citep{Buhlmann2011}, but cannot achieve the oracle property without further assumptions \citep{Zhao2006}.
The adaptive lasso \citep{Zou2006} or non-convex penalties \citep{Fan2001} can achieve the oracle property.

\section{Simulation}
\label{sec:simulation}

We conduct a simulation study to evaluate the performance of our method compared to existing integrative learning methods that do not incorporate graph information.
We compare fully heterogeneous (FHT; independent estimation and tuning) models, integrative homogeneity (IHM) models, and integrative heterogeneity (IHT) models.
IHM and IHT refer to the homogeneity model and the heterogeneity model, respectively, as defined in \citet{Zhao2015}.
We denote our SIL methods by SIL-Lasso, SIL-MCP, and SIL-LS, which use \eqref{eqn:Lasso}, \eqref{eqn:MCP}, and \eqref{eqn:logsum} for $\rho_1$, respectively.
The heterogeneity SIL-Lasso uses \eqref{eqn:Lasso} for $\rho_2$ while fixing $\alpha=1$ for its homogeneity version.
The homogeneity versions of SIL-MCP and SIL-LS use $\rho_2(\Gamma) = \| \Gamma \|_F$ and the heterogeneity versions use $\rho_2(\Gamma) = \| \Gamma^T \|_{2,1}$.

The FHT competing models include Lasso \citep{Tibshirani1996}, Enet \citep{Zou2005}, and SRIG \citep{Yu2016}, the IHM competitors include $L_2$ gMCP \citep{Liu2014Composite} and gLasso \citep{Ma2011}, and the IHT competitors include $L_1$ gMCP \citep{Liu2014Composite} and sgLasso (sparse gLasso), which uses
\begin{align*}
    P(B) = \lambda \alpha \|B^T\|_{2,1} + \lambda (1-\alpha) \|B\|_{1,1}.
\end{align*}

We describe how to generate the precision matrix of $X^m$.
For each $m = 1,\dots,M$, we generate a block diagonal matrix
\begin{align*}
\bs\Omega^m
=
\diag \left(
	\Omega_1^m, \dots, \Omega_B^m
\right),
\end{align*}
where each sub-matrix $\Omega_b^m$
is a $p_{B} \times p_{B}$ symmetric matrix.
We consider three different types of graphical structure for $\Omega_b^m$ depending on scenarios.
The detailed procedure goes as follows.
\begin{enumerate}
	\item Set $\Omega_b^m$ to a $p_{B} \times p_{B}$ zero matrix.
	
	\item Depending on scenarios, generate the nonzero lower triangular entries specified below as $\mc U \left( -1.5, -0.5 \right)$.
	
    \begin{itemize}
        \item Scenario 1 (ring type): $\left[ \Omega_b^m \right]_{p_B,1}$ and $\left[ \Omega_b^m \right]_{k,k-1}$ for $k>1$ are nonzero.
        
        \item Scenario 2 (hub type): $\left[ \Omega_b^m \right]_{k1}$ for $k>1$ are nonzero.

        \item Scenario 3 (random type): Each $\left[ \Omega_b^m \right]_{j,k}$ ($j>k$) is nonzero with probability $3/p_B$.
        
    \end{itemize}
    
    \item Fill in the upper triangular entries; $\Omega_b^m \leftarrow \Omega_b^m + \Omega_b^{mT}$.
    
    \item $\left[ \Omega_b^m \right]_{j j} \leftarrow 0.5 - \sum_{k=1}^{p_B} \left[ \Omega_b^m \right]_{j k}$.
    
    \item Normalize $\Omega_b^m$ such that the diagonal elements of its inverse matrix become 1.

\end{enumerate}

\begin{table}[!ht]
    \centering
    \scriptsize
    \caption{\label{tbl:simul_homo} Simulation results for homogeneity data. FHT; fully heterogeneous models, IHM; integrative homogeneity models, IHT; integrative heterogeneity models, $\mc G$; Y indicates the method incorporates graph information, MSE; mean squared prediction error, $L_2$; average $L_2$ distance between estimated coefficients and true coefficients, FPR; false positive rates, FNR; false negative rates.}
    \begin{tabular*}{\textwidth}{c@{\extracolsep{\fill}}lc|cccc|cccc|cccc}
\multirow{2}{*}{Type} & \multicolumn{1}{c}{\multirow{2}{*}{Method}} & \multirow{2}{*}{$\mc G$} & \multicolumn{4}{c}{Scenario 1} & \multicolumn{4}{c}{Scenario 2} &  \multicolumn{4}{c}{Scenario 3}\\
 &  &  & MSE & $L_2$ & FPR & FNR & MSE & $L_2$ & FPR & FNR & MSE & $L_2$ & FPR & FNR\\
 \hline
 \hline
\multirow{6}{*}{FHT} & \phantom{0}Lasso &  
& 1.274 & 1.509 & 0.330 & 0.238
& 1.348 & 1.902 & 0.402 & 0.234
& 1.274 & 1.539 & 0.313 & 0.297\\
&&& (.004) & (.010) & (.005) & (.005) & (.004) & (.014) & (.005) & (.006) & (.004) & (.014) & (.005) & (.008)\\
 & \phantom{0}Enet & 
& 1.274 & 1.509 & 0.330 & 0.238
& 1.348 & 1.902 & 0.402 & 0.234
& 1.274 & 1.539 & 0.313 & 0.297\\
&&& (.004) & (.010) & (.005) & (.005) & (.004) & (.014) & (.005) & (.006) & (.004) & (.014) & (.005) & (.008)\\
 & \phantom{0}SRIG & Y 
& 1.156 & 1.052 & 0.125 & 0.112
& 1.157 & 1.269 & 0.100 & 0.000
& 1.194 & 1.223 & 0.208 & 0.089\\
&&& (.004) & (.008) & (.003) & (.003) & (.003) & (.014) & (.005) & (.000) & (.005) & (.015) & (.006) & (.004)\\
 \hline
\multirow{10}{*}{IHM} & \phantom{0}gLasso &  
& 1.185 & 1.262 & 0.573 & 0.026
& 1.255 & 1.670 & 0.718 & 0.010
& 1.186 & 1.285 & 0.576 & 0.096\\
&&& (.004) & (.010) & (.008) & (.003) & (.004) & (.013) & (.008) & (.002) & (.004) & (.014) & (.009) & (.007)\\
 & \phantom{0}$L_2$ gMCP &  
& 1.157 & 1.116 & 0.156 & 0.231
& 1.164 & 1.097 & 0.152 & 0.056
& 1.144 & 1.070 & 0.161 & 0.254\\
&&& (.005) & (.019) & (.015) & (.015) & (.004) & (.017) & (.014) & (.007) & (.004) & (.016) & (.016) & (.012)\\
 & \phantom{0}SIL-Lasso & Y 
& 1.099 & 0.838 & 0.187 & 0.006
& 1.130 & 1.011 & 0.169 & 0.000
& 1.109 & 0.916 & 0.229 & 0.009\\
&&& (.004) & (.018) & (.019) & (.002) & (.003) & (.013) & (.016) & (.000) & (.004) & (.014) & (.017) & (.002)\\
 & \phantom{0}SIL-MCP & Y 
& 1.109 & 0.911 & 0.092 & 0.037
& 1.120 & 0.923 & 0.045 & 0.000
& 1.111 & 0.935 & 0.122 & 0.030\\
&&& (.004) & (.020) & (.015) & (.005) & (.003) & (.014) & (.009) & (.000) & (.003) & (.014) & (.014) & (.004)\\
 & \phantom{0}SIL-LS & Y 
& 1.102 & 0.864 & 0.119 & 0.028
& 1.120 & 0.921 & 0.066 & 0.000
& 1.107 & 0.912 & 0.115 & 0.033\\
&&& (.004) & (.018) & (.017) & (.004) & (.003) & (.014) & (.013) & (.000) & (.004) & (.013) & (.015) & (.005)\\
 \hline
\multirow{10}{*}{IHT} & \phantom{0}sgLasso &  
& 1.194 & 1.308 & 0.536 & 0.034
& 1.266 & 1.706 & 0.689 & 0.021
& 1.194 & 1.318 & 0.552 & 0.105\\
&&& (.004) & (.014) & (.016) & (.004) & (.004) & (.017) & (.013) & (.004) & (.004) & (.016) & (.015) & (.008)\\
 & \phantom{0}$L_1$ gMCP &  
& 1.190 & 1.260 & 0.052 & 0.348
& 1.190 & 1.140 & 0.065 & 0.136
& 1.170 & 1.164 & 0.068 & 0.345\\
&&& (.006) & (.017) & (.005) & (.009) & (.006) & (.026) & (.005) & (.012) & (.006) & (.020) & (.006) & (.011)\\
 & \phantom{0}SIL-Lasso & Y 
& 1.105 & 0.852 & 0.204 & 0.011
& 1.135 & 1.044 & 0.159 & 0.000
& 1.115 & 0.936 & 0.242 & 0.015\\
&&& (.004) & (.016) & (.019) & (.002) & (.003) & (.014) & (.014) & (.000) & (.004) & (.014) & (.018) & (.003)\\
 & \phantom{0}SIL-MCP & Y 
& 1.127 & 0.991 & 0.045 & 0.072
& 1.128 & 0.957 & 0.031 & 0.000
& 1.130 & 1.022 & 0.073 & 0.068\\
&&& (.004) & (.019) & (.007) & (.005) & (.005) & (.028) & (.006) & (.000) & (.004) & (.015) & (.009) & (.006)\\
 & \phantom{0}SIL-LS & Y 
& 1.116 & 0.923 & 0.061 & 0.060
& 1.125 & 0.944 & 0.046 & 0.000
& 1.123 & 0.974 & 0.084 & 0.064\\
&&& (.004) & (.018) & (.009) & (.004) & (.004) & (.015) & (.010) & (.000) & (.004) & (.014) & (.010) & (.006)\\
    \end{tabular*}
\end{table}

The true regression coefficients $\bs \beta^m$ is given by
\begin{align*}
\bs \beta^m =
\begin{bmatrix}
\bs\alpha^T \Omega_1^m &
\bs\alpha^T \Omega_2^m &
\mb 0 &
\cdots &
\mb 0
\end{bmatrix}^T,
\end{align*}
for some vector $\bs\alpha$.
To create heterogeneity, the second block of features is set to have no influence on the outcome variable with probability $p_{ht}$.
That is, we have
\begin{align*}
\bs \beta^m =
\begin{bmatrix}
\bs\alpha^T \Omega_1^m &
\mb 0 &
\mb 0 &
\cdots &
\mb 0
\end{bmatrix}^T \quad \textrm{w.p. } p_{ht}.
\end{align*}
For each scenario, each row of $X^m$ is independently sampled from
$\mc N \left( \mb 0, (\bs\Omega^m)^{-1} \right)$.
Then, the responses are generated by a linear model as follows.
\begin{align*}
\mathbf{y}^m
= X^m \bs \beta^m + \mathbf{e}^m,
\end{align*}
where $\mb e^m \sim \mc N \left( \mb 0, \sigma^{2}I \right)$.
We generate a total of $N=nM$ observations with each dataset assigned $n$ samples.

\begin{table}[!ht]
    \centering
    \scriptsize
    \caption{\label{tbl:simul_hetero} Simulation results for heterogeneity data. FHT; fully heterogeneous models, IHM; integrative homogeneity models, IHT; integrative heterogeneity models, $\mc G$; Y indicates the method incorporates graph information, MSE; mean squared prediction error, $L_2$; average $L_2$ distance between estimated coefficients and true coefficients, FPR; false positive rates, FNR; false negative rates.}
    \begin{tabular*}{\textwidth}{c@{\extracolsep{\fill}}lc|cccc|cccc|cccc}
\multirow{2}{*}{Type} & \multicolumn{1}{c}{\multirow{2}{*}{Method}} & \multirow{2}{*}{$\mc G$} & \multicolumn{4}{c}{Scenario 1} & \multicolumn{4}{c}{Scenario 2} &  \multicolumn{4}{c}{Scenario 3}\\
 &  &  & MSE & $L_2$ & FPR & FNR & MSE & $L_2$ & FPR & FNR & MSE & $L_2$ & FPR & FNR\\
 \hline
 \hline
\multirow{6}{*}{FHT} & \phantom{0}Lasso &  
& 1.245 & 1.427 & 0.295 & 0.247 & 1.310 & 1.821 & 0.357 & 0.257 & 1.244 & 1.458 & 0.278 & 0.313\\
&&& (.005) & (.011) & (.005) & (.005) & (.005) & (.015) & (.006) & (.007) & (.005) & (.015) & (.005) & (.008)\\
 & \phantom{0}Enet & 
& 1.245 & 1.427 & 0.295 & 0.247 & 1.310 & 1.821 & 0.357 & 0.257 & 1.244 & 1.458 & 0.278 & 0.313\\
&&& (.005) & (.011) & (.005) & (.005) & (.005) & (.015) & (.006) & (.007) & (.005) & (.015) & (.005) & (.008)\\
 & \phantom{0}SRIG & Y 
& 1.135 & 0.967 & 0.112 & 0.115 & 1.132 & 1.149 & 0.094 & 0.000 & 1.164 & 1.120 & 0.193 & 0.088\\
&&& (.004) & (.011) & (.003) & (.004) & (.004) & (.016) & (.006) & (.000) & (.005) & (.015) & (.006) & (.005)\\
 \hline
\multirow{10}{*}{IHM} & \phantom{0}gLasso &  
& 1.178 & 1.238 & 0.576 & 0.030 & 1.243 & 1.641 & 0.706 & 0.016 & 1.179 & 1.262 & 0.570 & 0.099\\
&&& (.004) & (.010) & (.008) & (.004) & (.004) & (.013) & (.008) & (.003) & (.004) & (.013) & (.009) & (.007)\\
 & \phantom{0}$L_2$ gMCP &  
& 1.159 & 1.123 & 0.157 & 0.265 & 1.169 & 1.152 & 0.209 & 0.080 & 1.147 & 1.090 & 0.180 & 0.281\\
&&& (.005) & (.021) & (.015) & (.016) & (.006) & (.024) & (.016) & (.011) & (.004) & (.018) & (.018) & (.015)\\
 & \phantom{0}SIL-Lasso & Y 
& 1.105 & 0.860 & 0.217 & 0.013 & 1.137 & 1.064 & 0.215 & 0.003 & 1.113 & 0.938 & 0.266 & 0.014\\
&&& (.005) & (.020) & (.019) & (.003) & (.004) & (.022) & (.020) & (.002) & (.004) & (.017) & (.018) & (.003)\\
 & \phantom{0}SIL-MCP & Y 
& 1.113 & 0.913 & 0.123 & 0.049 & 1.127 & 0.980 & 0.073 & 0.003 & 1.113 & 0.945 & 0.140 & 0.034\\
&&& (.004) & (.021) & (.015) & (.006) & (.005) & (.025) & (.010) & (.002) & (.004) & (.018) & (.014) & (.004)\\
 & \phantom{0}SIL-LS & Y 
& 1.107 & 0.884 & 0.137 & 0.039 & 1.126 & 0.970 & 0.085 & 0.003 & 1.110 & 0.926 & 0.156 & 0.033\\
&&& (.004) & (.020) & (.016) & (.005) & (.005) & (.025) & (.012) & (.002) & (.004) & (.017) & (.016) & (.005)\\
 \hline
\multirow{10}{*}{IHT} & \phantom{0}sgLasso &  
& 1.195 & 1.311 & 0.499 & 0.061 & 1.260 & 1.706 & 0.629 & 0.051 & 1.196 & 1.335 & 0.500 & 0.145\\
&&& (.004) & (.016) & (.019) & (.007) & (.005) & (.021) & (.016) & (.008) & (.005) & (.019) & (.019) & (.012)\\
 & \phantom{0}$L_1$ gMCP &  
& 1.191 & 1.246 & 0.073 & 0.369 & 1.192 & 1.197 & 0.079 & 0.192 & 1.166 & 1.164 & 0.071 & 0.381\\
&&& (.006) & (.020) & (.006) & (.012) & (.006) & (.029) & (.005) & (.013) & (.005) & (.023) & (.005) & (.012)\\
 & \phantom{0}SIL-Lasso & Y 
& 1.107 & 0.862 & 0.214 & 0.016 & 1.128 & 1.033 & 0.179 & 0.001 & 1.119 & 0.952 & 0.249 & 0.025\\
&&& (.005) & (.022) & (.018) & (.003) & (.004) & (.018) & (.017) & (.001) & (.006) & (.021) & (.018) & (.004)\\
 & \phantom{0}SIL-MCP & Y 
& 1.129 & 0.987 & 0.070 & 0.091 & 1.126 & 0.957 & 0.067 & 0.000 & 1.133 & 1.031 & 0.085 & 0.087\\
&&& (.004) & (.019) & (.007) & (.008) & (.005) & (.026) & (.006) & (.000) & (.005) & (.020) & (.008) & (.009)\\
 & \phantom{0}SIL-LS & Y 
& 1.117 & 0.913 & 0.082 & 0.079 & 1.124 & 0.970 & 0.073 & 0.001 & 1.122 & 0.976 & 0.096 & 0.080\\
&&& (.004) & (.020) & (.010) & (.006) & (.004) & (.020) & (.009) & (.001) & (.005) & (.020) & (.010) & (.008)\\
    \end{tabular*}
\end{table}

We consider $M=5$ datasets with $p=100$ features ($B=10$, $p_B=10$).
The error variance is $\sigma^2=1$ and we use $\bs\alpha = \begin{bmatrix} 1 & 1/3 & \cdots & 1/3 \end{bmatrix}^T$ for scenarios 1 and 3 and $\bs\alpha = \begin{bmatrix} 1 & 1/4 & \cdots & 1/4 \end{bmatrix}^T$ for scenario 2. This yields roughly 2.5 signal-to-noise ratio for all scenarios.
We use the validation method for tuning our methods.
The tuning parameters are selected simultaneously via a grid point search over the multi-dimensional tuning parameter space.
For example, IHM-SIL-LS for Scenario 1 in Table \ref{tbl:simul_homo} searches over the $25 \times 10 \times 6$ points of $(\lambda,\eta,\lambda_R)$.
The tuple that minimizes the validated prediction error was selected and used for predicting the testing data.
The training sample size is $n=200$, the validation sample size is $n_v = 200$, and the testing sample size is $n_t = 1000$.
Every method is fitted for a total of 100 replicates and tuned by the validation method.
In Tables, we report the simulation results evaluated by the mean squared prediction error (MSE), the average $L_2$ distance between the estimated coefficients and the true coefficients, the false positive rates (FPR), and the false negative rates (FNR).

In Table \ref{tbl:simul_homo}, we consider the case where all datasets have homogeneous sparsity structure, i.e., $p_{ht}=0$.
For all scenarios, the integrative approaches (IHM and IHT) tend to yield better performance than the fully heterogeneous methods (FHT), as the integrative approaches are able to take advantage of the common sparsity structure of  coefficients, except when the graph information is incorporated (SRIG).
Since our methods also uses the graphical knowledge, they have clearly improved performance than other existing integrative learning methods.
Particularly, our three IHM methods show the best performance among all.
Although our IHT versions seem to lose a little more weak signals than our IHM versions do, it is still much less serious than other existing IHT methods.
This demonstrates the advantages of incorporating network information into integrative learning.  

In Table \ref{tbl:simul_hetero}, some datasets can have different sparsity with $p_{ht} = 0.3$. 
We observe similar performance patterns as in Table \ref{tbl:simul_homo}.
Although all methods pose slightly worse FPR and FNR compared to Table \ref{tbl:simul_homo} due to the heterogeneity in sparsity structure of coefficients, our methods still have substantially improved variable selection performance over the ones with no graph incorporation or the non-integrative learning methods.
It is particularly worth noting that existing IHT methods show more worse results compared to the homogeneity data (Table \ref{tbl:simul_homo}) than our IHT methods do.
This again confirms the advantages of incorporating network information into integrative learning.  

As the proposed methods rely on the graph information, we conduct the sensitivity analysis taking into account uncertainty of the graphical knowledge and inconsistency with the regression coefficients.
In the sensitivity analysis, we randomly remove about 20\% of edges from the true graph and use the reduced graph as a working graph.
This mimics the intermediate situation where only strong interactions are known or the case where there are potentially missing edges (partial correlations) due to a screening of predictors.
In Table \ref{tbl:sensitivity}, we can see the performance of the methods that use the graph information deteriorates, while the methods that do not use the graph information remain similar compared to Table \ref{tbl:simul_homo}.
However, the difference for our methods is very small compared to that of SRIG, which seems attributed to the effect of integrative learning.
This lends support to robustness of our methods to misspecified graphical information with missing edges.

\begin{table}[!ht]
    \centering
    \scriptsize
    \caption{\label{tbl:sensitivity} Sensitivity analysis results. FHT; fully heterogeneous models, IHM; integrative homogeneity models, IHT; integrative heterogeneity models, $\mc G$; Y indicates the method incorporates graph information, MSE; mean squared prediction error, $L_2$; average $L_2$ distance between estimated coefficients and true coefficients, FPR; false positive rates, FNR; false negative rates.}
    \begin{tabular*}{\textwidth}{c@{\extracolsep{\fill}}lc|cccc|cccc|cccc}
\multirow{2}{*}{Type} & \multicolumn{1}{c}{\multirow{2}{*}{Method}} & \multirow{2}{*}{$\mc G$} & \multicolumn{4}{c}{Scenario 1} & \multicolumn{4}{c}{Scenario 2} &  \multicolumn{4}{c}{Scenario 3}\\
 &  &  & MSE & $L_2$ & FPR & FNR & MSE & $L_2$ & FPR & FNR & MSE & $L_2$ & FPR & FNR\\
 \hline
 \hline
\multirow{6}{*}{FHT} & \phantom{0}Lasso &  
& 1.267 & 1.508 & 0.322 & 0.250 & 1.349 & 1.925 & 0.400 & 0.226 & 1.279 & 1.565 & 0.329 & 0.285\\
&&& (.004) & (.009) & (.005) & (.005) & (.004) & (.014) & (.006) & (.006) & (.005) & (.014) & (.005) & (.007)\\
 & \phantom{0}Enet & 
& 1.267 & 1.508 & 0.322 & 0.250 & 1.349 & 1.925 & 0.400 & 0.226 & 1.279 & 1.565 & 0.329 & 0.285\\
&&& (.004) & (.009) & (.005) & (.005) & (.004) & (.014) & (.006) & (.006) & (.005) & (.014) & (.005) & (.007)\\
 & \phantom{0}SRIG & Y 
& 1.184 & 1.220 & 0.150 & 0.166 & 1.245 & 1.495 & 0.275 & 0.158 & 1.239 & 1.469 & 0.151 & 0.199\\
&&& (.005) & (.018) & (.006) & (.006) & (.006) & (.024) & (.009) & (.007) & (.006) & (.018) & (.005) & (.007)\\
 \hline
\multirow{10}{*}{IHM} & \phantom{0}gLasso &  
& 1.178 & 1.256 & 0.583 & 0.024 & 1.254 & 1.685 & 0.717 & 0.005 & 1.191 & 1.310 & 0.607 & 0.084\\
&&& (.004) & (.011) & (.008) & (.004) & (.004) & (.012) & (.007) & (.002) & (.004) & (.013) & (.009) & (.006)\\
 & \phantom{0}$L_2$ gMCP &  
& 1.148 & 1.080 & 0.251 & 0.166 & 1.161 & 1.108 & 0.168 & 0.042 & 1.146 & 1.080 & 0.182 & 0.214\\
&&& (.004) & (.019) & (.020) & (.016) & (.004) & (.017) & (.015) & (.006) & (.005) & (.019) & (.014) & (.010)\\
 & \phantom{0}SIL-Lasso & Y 
& 1.108 & 0.917 & 0.221 & 0.006 & 1.141 & 1.102 & 0.257 & 0.002 & 1.114 & 0.969 & 0.231 & 0.017\\
&&& (.004) & (.015) & (.020) & (.002) & (.003) & (.012) & (.016) & (.001) & (.004) & (.015) & (.017) & (.004)\\
 & \phantom{0}SIL-MCP & Y 
& 1.110 & 0.931 & 0.111 & 0.032 & 1.124 & 0.942 & 0.073 & 0.006 & 1.112 & 0.951 & 0.099 & 0.060\\
&&& (.004) & (.016) & (.017) & (.005) & (.004) & (.015) & (.010) & (.002) & (.004) & (.014) & (.015) & (.007)\\
 & \phantom{0}SIL-LS & Y 
& 1.106 & 0.906 & 0.124 & 0.030 & 1.122 & 0.949 & 0.082 & 0.008 & 1.107 & 0.927 & 0.081 & 0.054\\
&&& (.004) & (.015) & (.018) & (.005) & (.003) & (.011) & (.012) & (.002) & (.004) & (.013) & (.014) & (.006)\\
 \hline
\multirow{10}{*}{IHT} & \phantom{0}sgLasso &  
& 1.189 & 1.290 & 0.576 & 0.035
& 1.263 & 1.719 & 0.688 & 0.017
& 1.204 & 1.363 & 0.556 & 0.104\\
&&& (.004) & (.014) & (.016) & (.006) & (.004) & (.017) & (.012) & (.003) & (.005) & (.015) & (.016) & (.008)\\
 & \phantom{0}$L_1$ gMCP &  
& 1.187 & 1.254 & 0.070 & 0.350 & 1.179 & 1.124 & 0.056 & 0.132 & 1.164 & 1.141 & 0.074 & 0.321\\
&&& (.007) & (.020) & (.008) & (.011) & (.006) & (.023) & (.004) & (.012) & (.005) & (.020) & (.005) & (.012)\\
 & \phantom{0}SIL-Lasso & Y 
& 1.113 & 0.934 & 0.217 & 0.015 & 1.149 & 1.131 & 0.247 & 0.014 & 1.115 & 0.968 & 0.235 & 0.020\\
&&& (.004) & (.016) & (.019) & (.003) & (.004) & (.014) & (.015) & (.002) & (.004) & (.012) & (.017) & (.004)\\
 & \phantom{0}SIL-MCP & Y 
& 1.130 & 1.026 & 0.042 & 0.105 & 1.135 & 0.981 & 0.029 & 0.047 & 1.127 & 1.015 & 0.035 & 0.121\\
&&& (.004) & (.016) & (.006) & (.007) & (.004) & (.021) & (.004) & (.005) & (.004) & (.015) & (.005) & (.009)\\
 & \phantom{0}SIL-LS & Y 
& 1.120 & 0.971 & 0.058 & 0.090 & 1.137 & 0.997 & 0.049 & 0.060 & 1.122 & 0.989 & 0.044 & 0.110\\
&&& (.004) & (.015) & (.009) & (.006) & (.003) & (.013) & (.007) & (.004) & (.004) & (.016) & (.007) & (.008)\\
    \end{tabular*}
\end{table}

\section{Application}
\label{sec:application}

Alzheimer's disease (AD) is a major cause of dementia.
The Alzheimer's disease neuroimaging initiative (ADNI) is a large scale multisite longitudinal study where researchers at 63 sites track the progression of AD in the human brain through the process of normal aging, early mild cognitive impairment (EMCI), and late mild cognitive impairment (LMCI) to dementia or AD. Its goal is to validate diagnostic and prognostic biomarkers that can predict the progress of AD.

In our data analysis, we investigate the association of patients' gene expression levels with an imaging marker that captures AD progression. Specifically, we treat the fluorodeoxyglucose positron emission tomography (FDG-PET) averaged over the regions of interest (ROI) as the response variable, which measures cell metabolism. Cells affected by AD tend to show reduced metabolism. Since the association of FDG with gene expression levels may change at different stages of AD, we divide the total of 675 subjects into three groups depending on their baseline disease status, namely, CN (cognitively normal, $n=229$), MCI (EMCI+LMCI, $n=402$), and AD ($n=44$). 

The samples in each group are randomly split into a training set (50\%), a validation set (25\%), and a testing set (25\%).
For each split, we fit with our method and the existing methods considered in Section \ref{sec:simulation} plus some fully homogeneous models to check the heterogeneity of the datasets, and report the prediction errors for the testing samples. The regularization parameters of all methods are tuned by validation method and the graph information is obtained from KEGG. This procedure is repeated for 200 random splits of the data and the average squared prediction errors are reported in Table \ref{tbl:application}.

\begin{table}[!ht]
    \centering
    \scriptsize
    \caption{\label{tbl:application} Average prediction errors for ADNI dataset. FHM; fully homogeneous models, FHT; fully heterogeneous models, IHM; integrative homogeneity models, IHT; integrative heterogeneity models, $\mc G$; Y indicates the method incorporates graph information.}
    \begin{tabular*}{\textwidth}{c@{\extracolsep{\fill}}clcccccc}
 & Type & Method & $\mc G$ &  & MCI & AD & CN & \\
 \hline
 \hline
 & \multirow{3}{*}{FHM} & Lasso &  &  & 0.926 & 1.094 & 1.020\\
 &  & Enet &  &  & 0.901 & 1.075 & 0.987\\
 &  & SRIG & Y &  & 0.955 & 1.063 & 1.035\\
 \hline
 & \multirow{3}{*}{FHT} & Lasso &  &  & 0.916 & 1.028 & 0.996\\
 &  & Enet &  &  & 0.881 & 0.983 & 0.961\\
 &  & SRIG & Y &  & 0.933 & 1.035 & 1.008\\
 \hline
 & \multirow{5}{*}{IHM} & gLasso &  &  & 0.934 & 1.017 & 0.991\\
 &  & $L_2$ gMCP &  &  & 0.898 & 1.027 & 1.005\\
 &  & SIL-Lasso & Y &  & 0.873 & 0.946 & 0.946\\
 &  & SIL-MCP & Y &  & 0.876 & 0.947 & 0.948\\
 &  & SIL-LS & Y &  & 0.879 & 0.948 & 0.945\\
 \hline
 & \multirow{5}{*}{IHT} & sgLasso &  &  & 0.939 & 1.022 & 0.996\\
 &  & $L_1$ gMCP &  &  & 0.914 & 1.045 & 1.001\\
 &  & SIL-Lasso & Y &  & 0.862 & 0.950 & 0.940\\
 &  & SIL-MCP & Y &  & 0.878 & 0.934 & 0.955\\
 &  & SIL-LS & Y &  & 0.881 & 0.941 & 0.949\\
 \hline
    \end{tabular*}
\end{table}

\begin{table}[!ht]
    \centering
    \scriptsize
    \caption{\label{tbl:pathway} Ten enriched pathways and $p$-values for each method. '-' indicates not enriched in the genes selected by the method. FHM; fully homogeneous models, FHT; fully heterogeneous models, IHM; integrative homogeneity models, IHT; integrative heterogeneity models, $\mc G$; Y indicates the method incorporates graph information, P1; AGE-RAGE signaling pathway, P2; Angiopoietin receptor Tie2-mediated signaling, P3; Chemokine signaling pathway, P4; CXCR4-mediated signaling events, P5; Glucocorticoid receptor regulatory network, P6; IL2-mediated signaling events, P7; MAPKinase Signaling Pathway, P8; Prolactin signaling pathway, P9; Signaling by PDGF, P10; Tuberculosis.}
    \begin{tabular*}{\textwidth}{c@{\extracolsep{\fill}}clccccccccccccc}
 & Type & Method & $\mc G$ &  & P1 & P2 & P3 & P4 & P5 & P6 & P7 & P8 & P9 & P10 & \\
 \hline
 \hline
 & \multirow{3}{*}{FHM} & Lasso &  &  & - & - & - & - & - & - & - & - & - & -\\
 &  & Enet &  &  & - & - & - & - & - & - & - & - & - & -\\
 &  & SRIG & Y &  & - & - & - & - & - & - & - & - & - & -\\
 \hline
 & \multirow{3}{*}{FHT} & Lasso &  &  & - & - & - & - & - & - & - & - & - & -\\
 &  & Enet &  &  & - & - & - & - & - & - & - & - & - & -\\
 &  & SRIG & Y &  & 1.6e-5 & 5.8e-5 & - & - & 2.9e-4 & 7.3e-5 & 2.8e-4 & 1.8e-4 & - & - \\
 \hline
 & \multirow{5}{*}{IHM} & gLasso &  &  & - & - & - & - & - & - & - & - & - & -\\
 &  & $L_2$ gMCP &  &  & - & - & - & - & - & - & - & - & - & -\\
 &  & SIL-Lasso & Y &  & 2.1e-9 & 4.2e-6 & - & 9.7e-8 & 1.1e-6 & 5.8e-6 & 1.1e-6 & - & 1.6e-6 & 2.9e-6\\
 &  & SIL-MCP & Y &  & 1.1e-6 & 1.9e-6 & 2.1e-5 & 1.2e-6 & 1.6e-5 & 4.1e-8 & 1.6e-5 & 1.9e-7 & 4.9e-6 & 1.9e-5\\
 &  & SIL-LS & Y &  & 1.3e-6 & 1.1e-4 & - & - & 2.0e-5 & 1.5e-4 & 1.1e-4 & 1.0e-5 & 8.1e-5 & - \\
 \hline
 & \multirow{5}{*}{IHT} & sgLasso &  &  & - & - & - & - & - & - & - & - & - & -\\
 &  & $L_1$ gMCP &  &  & - & - & - & - & - & - & - & - & - & -\\
 &  & SIL-Lasso & Y &  & 5.8e-11 & 1.3e-9 & 7.7e-9 & 1.3e-12 & 1.3e-6 & 1.3e-7 & 1.3e-6 & - & - & 1.7e-7\\
 &  & SIL-MCP & Y &  & - & - & 2.1e-4 & 2.6e-4 & - & 1.5e-6 & - & 5.0e-6 & 2.7e-5 & 2.0e-4\\
 &  & SIL-LS & Y &  & 1.3e-6 & 1.1e-4 & - & 4.4e-5 & - & 1.5e-4 & 1.1e-4 & 1.0e-5 & - & -\\
 \hline
    \end{tabular*}
\end{table}

As shown in Table \ref{tbl:application}, all of the FHM methods tend to underperform the corresponding FHT methods, suggesting that the model of interest likely has different parameters for different groups.
Despite such heterogeneity, our methods show best prediction performance for all groups.
The existing integrative learning approaches, which do not incorporate network information, seem to have difficulty integrating information from different datasets.

Another benefit of incorporating graphical pathway information is enhanced interpretability of the selected genes.
To confirm, we conduct the pathway enrichment analysis based on the 30 most frequently selected genes of each method during the 200 repeats.
Table \ref{tbl:pathway} include 10 enriched pathways that are related to Alzheimer disease and the associated $p$-values.
Any method that does not incorporate graph information, including the existing integrative learning approaches, has no enriched pathway.
Except SILs, only the fully heterogeneous SRIG yields some enriched pathways.
However, the $p$-values of SRIG tend to be larger than those of our methods.

\section{Discussion}
\label{sec:discussion}

We have proposed a novel integrative learning method, called SIL, which can incorporate the graphical structure of features.
SIL possesses appealing theoretical properties, is scalable to high-dimensional data, and has been shown to outperform existing integrative learning methods through a simulation study and a real data analysis. 
 
In practice, the ground truth sparsity structure of $\bs\beta^0$ may not be consistent with the graphical structure.
However, when the discrepancy is moderate, our proposed method will still show reasonably good performance by detecting the subset of groups that cover all or most of the nonzero coefficients.
Note that the sensitivity analysis (Table \ref{tbl:sensitivity}), which is conducted partly in consideration of such inconsistency, suggests the proposed method is quite robust.
Even when the graphical information is completely irrelevant to the sparsity structure, our method will not fail.
The tuning procedure will discourage the group-wise selection and we can expect the performance to be comparable to that of the plain ridge regression.

On the other hand, it is widely acknowledged that the graph information obtained from existing databases could be inaccurate or incomplete.
It is potentially of future interest to investigate approaches that are more robust to incomplete graph information.
One potential approach is to combine the graph information from existing databases and the estimated graph information using the data being analyzed.
Another direction for future research is to incorporate graph information that may vary between datasets.

\section*{Acknowledgements}

This work is partly supported by NIH grant RF1AG063481.
The content is solely the responsibility of the authors and does not necessarily represent the official views of the National Institutes of Health.
The complete ADNI Acknowledgement is available here  (\href{http://adni.loni.usc.edu/wp-content/uploads/how\_to\_apply/ADNI\_Acknowledgement\_List.pdf}{click}).

\bibliographystyle{apalike}
\bibliography{SIL}  






\end{document}

%% file: MyShort.tex
\newcommand\mb[1]{\mathbf{#1}}
\newcommand\mc[1]{\mathcal{#1}}

\newcommand\bb[1]{\mathbb{#1}}
\newcommand\bs[1]{\boldsymbol{#1}}

\newcommand\whbs[1]{\widehat{\boldsymbol{#1}}}

\def\diag{\qopname\relax o{diag}}
\def\vect{\qopname\relax o{vec}}

\def\argmin{\qopname\relax m{argmin}}

\def\diag{\qopname\relax o{diag}}
\def\sign{\qopname\relax o{sign}}
\def\supp{\qopname\relax o{supp}}

\def\prox{\qopname\relax o{prox}}

\newtheorem{ass}{Assumption}
\newtheorem{thm}{Theorem}

\newtheorem{pro}{Proposition}
\newtheorem{cor}{Corollary}
\newtheorem{rem}{Remark}